\documentclass[prd,twocolumn,citeautoscript,superscriptaddress,notitlepage,longbibliography,bibnotes,footinbib]{revtex4-1}   
\pdfoutput=1  
\usepackage{amsmath,amssymb,mathrsfs,bm,feynmf,setspace,xspace,soul,empheq}  
\usepackage{graphicx,tikz,xparse}         
\usepackage[tight]{subfigure}           
\usepackage{color} 
\usepackage[colorlinks=true]{hyperref}         
\hypersetup{
    bookmarks=true,         % show bookmarks bar?
    unicode=false,          % non-Latin characters 
    pdftoolbar=true,        % show Acrobat
    pdfmenubar=true,        % show Acrobat 
    pdffitwindow=false,     % window fit to page when opened
    pdfstartview={FitH},    % fits the width of the page to the window
    pdftitle={My title},    % title
    pdfauthor={Author},     % author
    pdfsubject={Subject},   % subject of the document
    pdfcreator={Creator},   % creator of the document
    pdfproducer={Producer}, % producer of the document
    pdfkeywords={keyword1} {key2} {key3}, % list of keywords
    pdfnewwindow=true,      % links in new window
    colorlinks=true,       % false: boxed links; true: colored links
    linkcolor=magenta, %red,          % color of internal links (change box color with linkbordercolor)
    citecolor=blue,        % color of links to bibliography
    filecolor=magenta,      % color of file links
    urlcolor=cyan           % color of external links
} 
 
\newcommand{\bra}[1]{\left\langle #1\right|}
\newcommand{\ket}[1]{\left| #1\right\rangle}

\newcommand{\ba}{\begin{array}{ccc}}
\newcommand{\ea}{\end{array}}

\def\bea{\begin{eqnarray}}
\def\eea{\end{eqnarray}}
\newcommand{\bml}{\begin{multline}}

\newcommand{\eeqm}{\end{multline}}

\newcommand{\bsp}{\begin{split}}
\newcommand{\esp}{\end{split}}

\newcommand{\mc}{\mathcal}

\newcommand{\req}[1]{Eq.\thinspace(\ref{eq:#1})}

\newcommand{\ts}{\thinspace{}}
\newcommand{\rfig}[1]{Fig.\ts\ref{fig:#1}}

\DeclareMathOperator{\Tr}{Tr}

\newcommand{\bl}[1]{{\color{black}#1}}  

\begin{document} 

\title{Quantum spin chains with multiple dynamics} %critical

\author{Xiao Chen}
%\email{xchen@kitp.ucsb.edu}
\affiliation{Kavli Institute for Theoretical Physics,
University of California at Santa Barbara, CA 93106, USA}

\author{Eduardo Fradkin}   
%\email{efradkin@illinois.edu} 
\affiliation{Department of Physics and Institute for Condensed Matter Theory, University of Illinois at Urbana-Champaign, 1110 West Green Street,
Urbana, Illinois 61801-3080, USA}  

\author{William Witczak-Krempa}  
%\email{w.witczak-krempa@umontreal.ca} 
\affiliation{D\'epartement de physique, %and Regroupement qu\'eb\'equois sur les mat\'eriaux de pointe, 
Universit\'e de Montr\'eal, Montr\'eal (Qu\'ebec), H3C 3J7, Canada} 

 \date{\today}   

\begin{abstract} 
Many-body systems with multiple emergent time scales arise in various contexts, including classical
critical systems, correlated quantum materials, and ultra-cold atoms.
We investigate such non-trivial  
quantum dynamics in a new setting: a spin-1 bilinear-biquadratic chain. It has a solvable entangled groundstate, but a 
gapless excitation spectrum that is poorly understood. 
By using large-scale DMRG simulations, we find that the lowest excitations have a dynamical exponent $z$ that varies from 
2 to 3.2 as we {\color{black}vary} a coupling in the Hamiltonian. We find an additional 
gapless mode with a continuously varying exponent $2\leq z <2.7$, which establishes the presence of multiple dynamics.
In order to explain these striking properties, {\color{black}we construct a continuum wavefunction for the groundstate, which 
correctly describes the correlations and entanglement properties. 
We also give a continuum parent Hamiltonian, but show that additional ingredients are needed to capture the excitations of the chain.
%but fails to explain the excitations. 
By using an exact mapping to the \emph{non-equilibrium} dynamics of a classical spin chain, we find that the large dynamical exponent is due to subdiffusive spin motion.}  
Finally, we discuss the connections to other spin chains and to a family of quantum critical models in 2d.    
\end{abstract}  
 
\maketitle    

It is common for many-body quantum systems to possess multiple time-scales that determine the low-energy dynamics.
In a gapless system, the dynamics will be characterized by the dispersion relation of  
the excited states (quasiparticles need not exist), $E=A k^z$, where $k$ is the wavevector of the mode and $z$ the dynamical exponent. Different modes can have different $z$ exponents.  
For instance, a metal near a quantum critical point can have different dispersions for the electrons and the various 
order parameter fluctuations \cite{Hertz76,Millis93,ssbook,Oganesyan-2001,Meng2012,Lederer-2016}. 
However, this phenomenon has been far less studied in other types of systems.
Many studies have examined simpler systems, such as models described by relativistic conformal field theories having $z=1$, which enjoy additional symmetries that constrain the dynamics \cite{yellow,ssbook,Lucas17}.     
%In general, the $z$ exponents need not be integers. Such systems are often challenging to study, especially above one dimension. 

In this work, we reveal multiple dynamical exponents in a new setting: a strongly correlated 1d spin system. %with multiple dynamical exponents. 
Further, these exponents will be shown to   
vary continuously as a function of a coupling in the Hamiltonian. The spin $1$ quantum spin chain in question is a generalization of 
the so-called Motzkin Hamiltonian introduced by Bravyi \emph{et al} \cite{bravyi}. Its groundstate can be determined exactly but
not its excitation spectrum. With the help of large-scale Density Matrix Renormalization Group (DMRG) simulations, we discover
low lying excitations with different dynamical exponents. In order to gain insight in the low-lying spectrum we determine
a continuum version of the groundstate, and find a parent Hamiltonian. {\color{black}The latter possesses an excitation spectrum 
that is distinct from the spin chain but can provide useful insight into the construction
of the full low energy field theory.} 
This illustrates how a given groundstate can have starkly different excitations, and 
offers some guidance in the construction of the true low-energy description of the chain. Owing to the Rokhsar-Kivelson \cite{Rokhsar-1988} (RK)
structure of the spin Hamiltonian, we are able to connect the problem of determining the excitation spectrum to
studying the \emph{non-equilibrium} dynamics of the corresponding classical 1d chain \cite{Henley04}. This sheds light on the 
subdiffusive nature ($z>2$) of the excitations observed with DMRG. Finally, we provide connections to a family of two dimensional quantum 
critical system that have conformally invariant wavefunctions \cite{Ardonne-2004,Isakov2011}.    

\begin{figure}
\centering
\includegraphics[width=2.2in]{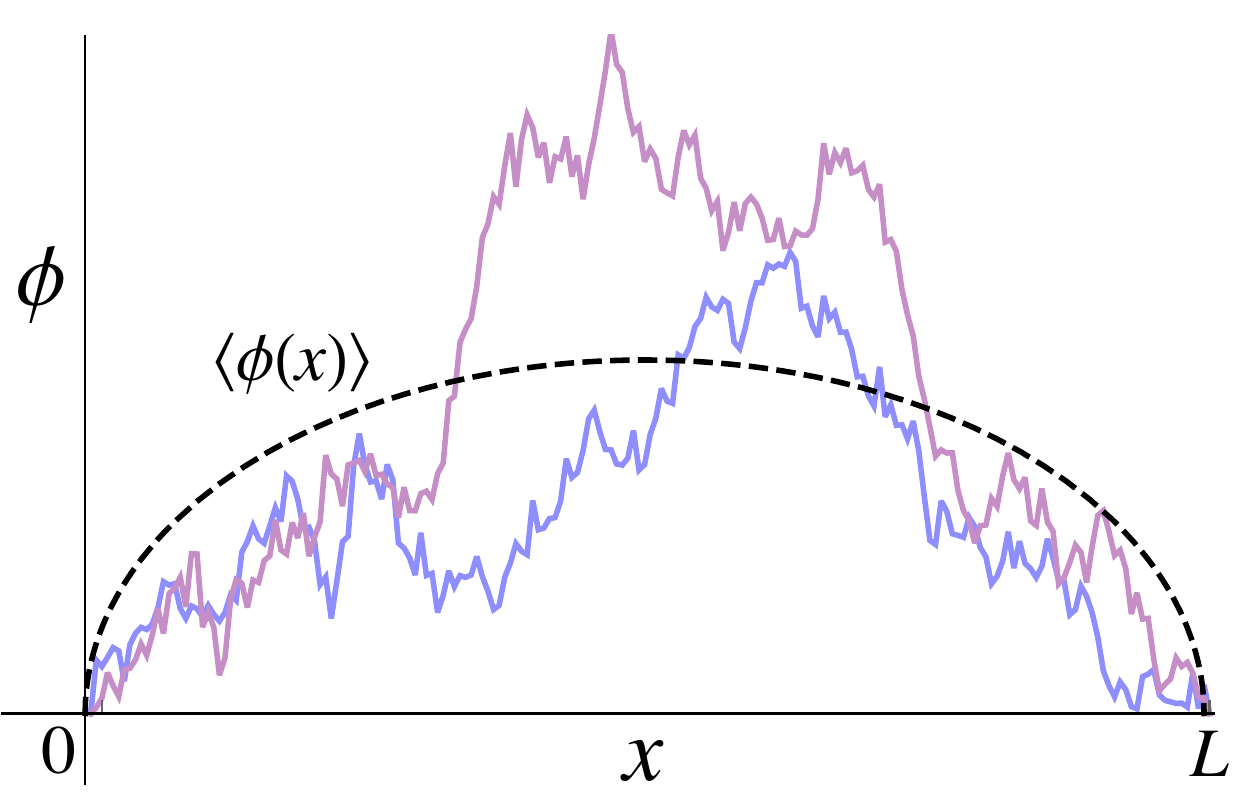} 
\caption{Representation of 2 Motzkin paths via the height variable $\phi$. Each path can be interpreted as a Brownian excursion.
 The dashed line is the average of $\phi$ in the groundstate \eqref{psi0}.  
The spin $\langle S^z\rangle\!=\!\langle \partial_x\phi \rangle$ tends to zero deep in the bulk.}  
\label{fig:excursion}  
\end{figure}

{\bf Critical quantum spin chain.} The Hamiltonian describes $N$ $S\!=\! 1$ spins interacting via nearest neighbor exchange:  
\begin{align} \label{Hbulk}
  H_{\rm bulk}\! =\! \sum_{i=1}^{N-1} \ket{D}_{\!i,i+1}\!\!\bra{D} + \ket{U}_{\!i,i+1}\!\!\bra{U}+ c \ket{V}_{\!i,i+1}\!\!\bra{V}
\end{align}
where $\ket{D}\!\sqrt 2\!=\! \ket{0d}-\ket{d0}$, $\ket{U}\!\sqrt 2\!=\! \ket{0u} -\ket{u0}$, 
$\ket{V}\!\sqrt 2\!=\! \ket{00} -\ket{ud}$; $c\geq0$ is a free parameter. Here, $u,d,0$ %$\{u\!=\!\mathrm{up},d\!=\!\mathrm{down},0\}$ 
label the $S^z$ eigenstates.  
In terms of the spin operators $S^{x,y,z}$, Eq.\eqref{Hbulk} takes the form of an anisotropic bilinear-biquadratic Hamiltonian 
$\sum_i(A_{ab}S_i^a S_{i+1}^b + B_{abcd}S_i^a S_i^b S_{i+1}^c S_{i+1}^d$); 
we give the coefficients $A,B$ in Supplemental Material \cite{SM}.   %Appendix~\ref{ap:spin}.   
We will work with open chains with an additional boundary term% acting on sites $1,N$: 
\begin{align} \label{H}
  H = H_{\rm bulk} + \tfrac12 S_1^z(S_1^z -1)+\tfrac12 S_N^z(S_N^z+1)
\end{align}
%$H_{\rm bdy}=\tfrac12 S_1^z(S_1^z -1)+\tfrac12 S_N^z(S_N^z+1)$. 
$H$ has a global U(1) symmetry generated by $S_{\rm tot}^z=\sum_i S_i^z$ \cite{Movassagh2017}.  
When $c=1$, $H$ reduces
to the so-called Motzkin Hamiltonian \cite{bravyi}. In that case, the groundstate 
is the equal weight superposition of all states corresponding to Motzkin paths. 
% the Motzkin state $\ket{\mc M_N}$, which  
For $N\!=\!3$:
\begin{align}
\label{motzkin}
   \ket{\mc M_3} = \tfrac{1}{\sqrt 4}( 
|\begin{tikzpicture}
\draw[gray, thick] (-.32,0) -- (-.04,0); 
\draw[gray, thick] (0,0) -- (.28,0); 
\draw[gray, thick] (.32,0) -- (.6,0);
\end{tikzpicture}
\rangle 
+
|\begin{tikzpicture}
\draw[gray, thick] (-.32,0) -- (-.04,0); 
\draw[gray, thick] (0,0) -- (.2,.2); 
\draw[gray, thick] (.24,.2) -- (.44,0);
\end{tikzpicture}
\rangle 
+
|\begin{tikzpicture}
\draw[gray, thick] (0,0) -- (.2,.2); 
\draw[gray, thick] (.24,.2) -- (.44,0);
\draw[gray, thick] (.48,0) -- (.76,0); 
\end{tikzpicture}
\rangle 
+
|\begin{tikzpicture}
\draw[gray, thick] (0,0) -- (.2,.2); 
\draw[gray, thick] (.24,.2) -- (.52,.2);
\draw[gray, thick] (.56,.2) -- (.76,0); 
\end{tikzpicture}
\rangle 
)
\end{align}
with the notation $u=\begin{tikzpicture}
\draw[gray, thick] (0,0) -- (.2,.2);\end{tikzpicture}$, $d=\begin{tikzpicture}
\draw[gray, thick] (0,.2) -- (.2,0);\end{tikzpicture}$, $0=\begin{tikzpicture}
\draw[gray, thick] (0,.1) -- (.28,.1);\end{tikzpicture}$. 
This allows for the height representation \cite{bravyi, RamisShor16, Movassagh2017, Zhang_Klich_2016} shown in \eqref{motzkin} and in \rfig{excursion}: the height variable $\phi_i$
is pinned to zero at both ends, $\phi_0=\phi_N=0$, while for $i\geq 1$ we have 
%\begin{align} \label{phi-def}
 $ S_i^z = \phi_i-\phi_{i-1}$.
%\end{align}
In this language, a Motzkin path has $\phi_i\geq0$ while being pinned to zero at the extremities.     
% At $N=4$ there are 9 possible Motzkin paths. For instance, the 9-state superposition for $|\mathcal M_4\rangle$ contains 
% the following maximal-height state  
% $\left|\begin{tikzpicture}
% \draw[gray, thick] (0,0) -- (.1,.1); 
% \draw[gray, thick] (.12,.12) -- (.22,.22);
% \draw[gray, thick] (.25,.22) -- (.35,.12);
% \draw[gray, thick] (.37,.10) -- (.47,0);
% \end{tikzpicture}
% \right\rangle$.
%$\ket{{\mathrlap{\diagup} } \,{\color{white}1} \,\,{\mathrlap{\diagdown} } \,{\color{white}1} \, \,}$ 
By virtue of being an equal weight superposition, the Motzkin groundstate $|\mathcal M_N\rangle$ is annihilated by  
all 3 projectors in Eq.\eqref{Hbulk}. It is thus a groundstate when $c\geq 0$. 
At $c=1$, it was shown to be the unique groundstate
as a consequence of the boundary term, a fact which remains true as long as $c>0$ \cite{bravyi}.  
At the special point $c=0$, other states belong to the groundstate manifold such as the all-0 product state. 
%%%%%%%%%%%%
%%%%%%%%%%%%

{\bf Groundstate in the continuum.} 
In the continuum limit where $x$ spans distances much larger than the lattice spacing, 
the groundstate wavefunction of the Motzkin Hamiltonian \eqref{H} takes the simple form
\begin{align} \label{psi0}
\Psi_0[\phi(x)]=\frac{1}{\sqrt Z} e^{-\tfrac{\kappa}{2} \int_0^L\! dx\, (\partial_x\phi)^2}\prod_x\theta(\phi(x)) 
\end{align}
which is defined in terms of the (coarse-grained) height field $\phi$ introduced above. 
This is reminiscent of the wavefunction of the quantum Lifshitz model in 2d \cite{Ardonne-2004,Fradkin-book}, 
with the distinction that $\phi$ here is non-compact. We discuss further connections between these models in the Outlook.
To match the boundary conditions of the lattice 
wavefunction, we impose the Dirichlet condition $\phi(0)=\phi(L)=0$ for a chain of length $L$. 
In this language, the spin field is given in the continuum by %analogue of \eqref{phi-def}, 
$S^z=d\phi/dx$;   
$\kappa$ is a parameter whose value will be fixed later and $\theta(\phi)$ is the Heaviside function that enforces $\phi$ 
to be non-negative. This constraint is necessary to obtain the Motzkin state (see Eq.\eqref{motzkin}).  
The normalization factor $Z$ takes the form of a (0+1)-dimensional partition function:
\begin{align}
  Z = \int_{\!\phi(0)=\phi(L)=0} \!\!\!\mc D\phi(x)\; e^{- \kappa \int_0^L \! dx (\partial_x\phi)^2 } \prod_x\theta(\phi(x))
\label{eq:Z_par}
\end{align}  
The exponential term in the wavefunction \eqref{psi0}, which determines the probability of a path $\phi(x)$, can be understood by mapping 
the problem to a random walk \cite{bravyi}. Let us momentarily go back to the lattice, which means
that we need to consider discrete Brownian motion in 1d restricted to the \emph{non-negative} integers $\phi_i\geq 0$. 
Taking the horizontal axis of the path $i$ 
as the time direction, the random walk can be illustrated as follows. 
%At each step where $\phi_i>0$, 
The walker takes a step chosen out of the 3 options: 
1) move up by one, 
2) move down by one, 
3) stay at the same place.  
The walk is subject to the constraint $\phi_i \geq 0$, and it must start/end at the same point, $\phi=0$, but is otherwise random.
% If $\phi_i=0$, only the ``up'' and ``stay'' steps are possible.
% There is a global constraint that enforces the walker to start and finish at the same point, $\phi=0$.
This process is called a Brownian excursion, and is illustrated in \rfig{excursion}. 
Any valid path constructed out of a succession of such steps has the same probability, whose value is given by the Motzkin wavefunction squared $P[\phi_i]=|\langle\phi_i|\mc M_N\rangle|^2$. $P$ thus equals the inverse
of the total number of Motzkin paths.  
Taking the long time limit, the random walk is described by a Langevin equation for the continuum field $\phi(x)\geq 0$.  
Statistical physics \cite{Majumdar2005} then tells us that the probability of a given path is given by the amplitude squared $|\Psi_0[\phi(x)]|^2$ 
of our wavefunction \eqref{psi0}.           
For the Moztkin type random walk, the variance at a typical step is $\sigma^2\!=\!\tfrac13(1^2+0^2+(-1)^2)=2/3$. 
This determines the diffusion constant in the long time limit, $1/(4\kappa)=\sigma^2/2=1/3$, i.e.\ $\kappa\!=\! 3/4$.  
Equipped with our parameter-free wavefunction, we can compute properties of the groundstate in the continuum limit. {\color{black}By comparing these properties with that for the discrete groundstate, we can show that Eq.\eqref{psi0} and the groundstate for Eq.\eqref{H} are exactly the same in the thermodynamical limit.}
For instance, the expectation value of the spin is $\langle S^z(x) \rangle\!=\!\langle \partial_x\phi \rangle =  (L-2x)/\sqrt{\pi\kappa L(L-x)x}$, 
which changes sign going from the left to the right end, see \rfig{excursion}. The non-zero expectation value arises due to 
the boundary conditions. 
Indeed, deep in the bulk $\langle S^z(\tfrac{L}{2}+a)\rangle \propto a/L^{3/2}$ rapidly vanishes as $L\to \infty$ at fixed $a$. 
This matches the calculation using the lattice wavefunction \cite{bravyi,Movassagh2017, Dellanna16}.           

The Motzkin groundstate is highly entangled in the sense that the R\'enyi entanglement entropy (EE) has a logarithmic scaling with subsystem size \cite{bravyi,Movassagh2017}.       
By considering the subregion $A$ to be the interval $[0,L_A]$, we find using \eqref{psi0}:  
\begin{align} \label{EE}
  S_n = \frac12 \ln\left( \frac{L_A(L-L_A)}{\epsilon L} \right) + b(n)
\end{align}
where the logarithm's prefactor is independent of the R\'enyi index $n$ and of $\kappa$; $\epsilon$ is a short-distance cutoff. 
The constant $b(n)$ depends on $n,\kappa$, and when we fix $\kappa=3/4$ we find an
exact agreement with the lattice calculation \cite{bravyi,Movassagh2017}. 
The calculation of Eq.\eqref{EE} is greatly simplified by the special form of the wavefunction Eq.\eqref{psi0}, allowing us to adapt
the methods of \cite{Fradkin06}, described in the Supplemental Material \cite{SM}. %(Appendix \ref{ap:qft}). 
Although in the limit $L_A\ll L$, the EE scales as $\frac12\ln L_A$, the complete form of the EE is distinct from what is found in CFTs, 
and implies that the long-distance limit
of the chain is not described by a CFT \cite{bravyi}. If we take region $A$ to be an interval located deep inside the bulk, 
we find $S_n=\tfrac12 \ln \tfrac{L_A}{\epsilon}$. 

The Motzkin wavefunction shows other clear differences from the groundstate of a
CFT, and is in fact less entangled. This can be seen by studying the mutual information 
for 2 disjoint intervals $A,B$, {\color{black}which was not studied before}.  The mutual information is defined as 
$I(A,B)= S(A)+S(B)-S(A\cup B)$.   
It measures the quantum correlations between $A$ and $B$,
giving an upper bound for two-point correlation functions of local observables \cite{Wolf2008}.  
In the wavefunction \eqref{psi0}, for 2 disjoint intervals deep inside the bulk, we find $I(A,B)\!=\! 0\bl{+O(L_A L_B/L^2)}$ \cite{SM}.
%(see Supplemental Material Appendix~\ref{ap:qft}).  
This is consistent with the result for the spin 2-point function:  %$\langle S^z(x_1)S^z(x_2)\rangle=\langle\partial_x\phi(x_1)\partial_x\phi(x_2)\rangle=0$
$\langle S^z(x_1)S^z(x_2)\rangle =0$ if $x_1\neq x_2$
in the Motzkin wavefunction \cite{Movassagh2017}, which can be readily derived using our continuum wavefunction \cite{long-prep}. %(Appendix~\ref{ap:qft}). 
The vanishing of $I(A,B)$ can be understood by using the above mapping between the wavefunction \eqref{psi0} and the random walk problem. 
Deep inside the bulk, we can ignore the boundary conditions and remove the constraint $\phi>0$  
due to the exponentially small probability for $\phi$ being near zero in \eqref{psi0}. % (without Heaviside functions). 
In this regime the random walk reduces to regular Brownian motion, instead of the constrained Brownian excursion.
Therefore, the probability for a walker moving a distance $\delta \phi$ in ``time'' $\delta x$ is independent of the history. 
There are essentially no correlations between the 2 disjoint intervals 
and we expect that the mutual information between them in the quantum state vanishes. 
This stands in contrast with CFTs, for which the mutual information between two well-separated small intervals with distance $r$ scales as $1/r^{\Delta}$, with $\Delta$ determined by the scaling dimension of primary operators \cite{Calabrese-2010}.

% Using the property that the two disjoint intervals deep in the bulk are essentially uncorrelated, we can also argue 
% that the entanglement negativity vanishes. 
% We now turn to the logarithmic negativity, which detects the entanglement between 2 disjoint
% regions and is 
% defined as $\mc N\!=\! \ln \|\rho_{A\cup B}^{T_B}\|$, 
% where $\rho_{A\cup B}^{T_B}$ denotes the partial transpose of the reduced density matrix $\rho_{A\cup B}$ with respect to region $B$. 
% $\|\mc O \|$ is the sum of the absolute value of the eigenvalues of $\mc O$.  
% For 2 intervals deep in the bulk, 
% $\rho_{A\cup B}\!=\! \int\! \mathcal D \phi_A\int\! \mathcal D \phi_B f(\phi_A,\phi_B)|\phi_A,\phi_B\rangle\langle \phi_A,\phi_B|$
% is diagonal and invariant under the partial transpose, in contrast to the generic case.   
% Thus $\mc N \!=\!\ln\mathrm{tr}(\rho_{A\cup B}) \!=\! 0$, further supporting the claim that the Motzkin wavefunction  
% is less entangled than the groundstate of a CFT \cite{Calabrese2012}.  

The fact that the EE of a single interval diverges logarithmically but the mutual information between 2 intervals
vanishes, suggests that non-local degrees of freedom are responsible for the entanglement measured via the EE. These escape
the more ``local'' 2-interval measures. 

{\bf A field theory with the Motzkin groundstate.} We now take a step further and construct \bl{one} quantum field theory whose groundstate 
is \eqref{psi0}, and has $z\neq 1$. The Hamiltonian of the field theory reads  
\begin{align} \label{orbi}
  H_{\rm orb} = \int dx \left( \frac12 \Pi^2 +\frac{\kappa^2}{2}(\partial_x^2\phi)^2 +V(\phi) \right)
\end{align} 
$\Pi$ is the canonical conjugate to the height operator $\phi$.
$V(\phi)$ is the potential that enforces the constraint $\phi\geq 0$: $V(\phi\!<\!0)\!=\!\infty$ and is zero otherwise. 
Thus, the target space of $\phi$ is the positive half-line, i.e.\ the orbifold  
obtained by moding the real line by the transformation $\phi\to -\phi$.  
To show that \eqref{psi0} is the groundstate 
of $H_{\rm orb}$, we can rewrite the latter as  
$H_{\rm orb}\!=\!\int_x(Q^\dag(x) Q(x)+V(\phi))$, where we have subtracted an infinite groundstate energy
and defined the annihilation operator \cite{Ardonne-2004,Fradkin-book} $Q(x)=\tfrac{1}{\sqrt 2}(\tfrac{\delta}{\delta\phi} -\kappa \partial_x^2\phi)$. 
The groundstate of Eq.\eqref{orbi} is annihilated by $Q$. This defines a functional equation that is satisfied 
by Eq.\eqref{psi0}: $Q\Psi_0[\phi]=0$.      

 Since $\Pi=\partial_t \phi$, we see that  
Eq.\eqref{orbi} is invariant under the spacetime dilation $x\to \lambda x$ and $t\to\lambda^2 t$ (with an appropriate field rescaling), implying that 
this Hamiltonian has dynamical exponent $z=2$ and is thus not a CFT, in agreement with the EE results above. 
Now, $H_{\rm orb}$ and the continuum limit of the Motzkin Hamiltonian share the same groundstate, but do they have the same low energy excitations? 
To answer this question, we now investigate the excited states of Motzkin Hamiltonian Eq.\eqref{H}. 
Because the problem is not readily amenable to  analytical calculations, we turn to DMRG simulations. 

\begin{figure}
\centering
\includegraphics[width=2.75in]{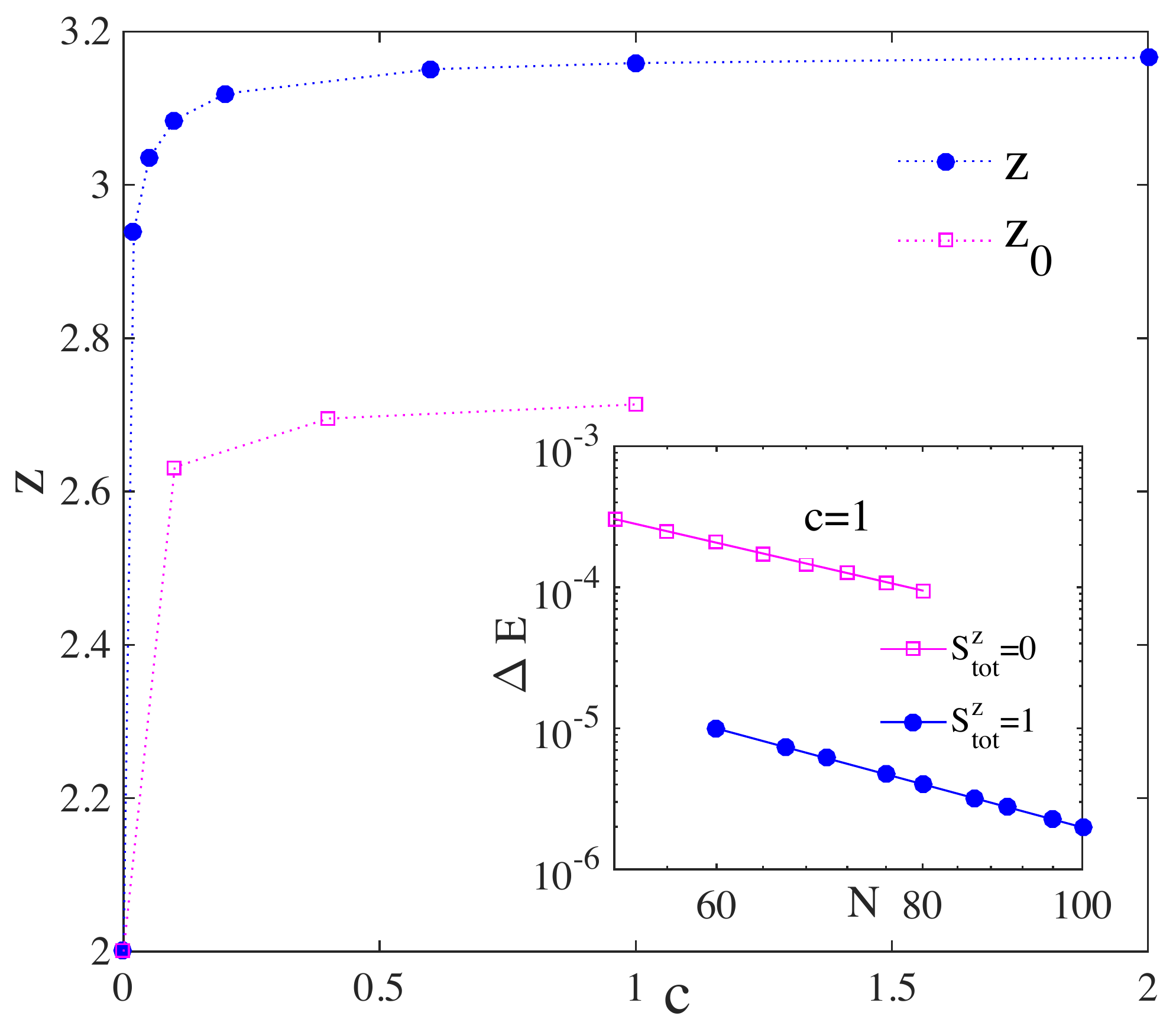}
\caption{DMRG data for the dynamical exponent of the generalized Motzkin Hamiltonian Eq.\eqref{H} versus the coupling $c$. The blue circles give $z$ for the lowest excitation, which has $S^z_{\rm tot}=1$. The pink squares give $z_0$ for the lowest excitation in the $S^z_{\rm tot}=0$ sector. {\bf Inset}: log-log plot of the energy
gap versus system size $N$ used to extract $z$ via $\Delta E\propto 1/N^z$.} 
\label{fig:z} 
\end{figure}

{\bf DMRG \& dynamical exponents.}  
At $c=1$, the many-body gap was shown to scale as $1/N^z$,  
with the analytical bound $z\geq 2$ \cite{RamisShor16,Gosset16},   
suggesting that our above field theory is a viable candidate to describe the Motzkin Hamiltonian. 
However, exact diagonalization (ED) \cite{bravyi} on small systems yielded $z\!=\!2.91$, while
previous DMRG calculations \cite{Dellanna16} yielded $z=2.7\pm0.1$. As we shall see, the former result suffers from strong
finite size effects, while the latter does not correspond to the true lowest excited state.  
In order to understand the spectrum, we have performed large-scale DMRG calculations using the ITensor library,
which we benchmarked using ED for short chains \cite{SM}.  
%Details of the simulations  are given in \cite{SM}. %(Appendix~\ref{ap:dmrg}).  
The results for $z$ as a function of $c$ are shown in \rfig{z}. At $c=1$, we find $z=3.16$ by using chains of length up to $N=100$,
an exponent substantially larger than the numerical results quoted above.
The excitation associated with this dynamical exponent is two-fold degenerate, with the 2 states having quantum numbers $S_{\rm tot}^z=\pm 1$, respectively. 
Interestingly, we also found a singly-degenerate excited state with higher energy in the $S_{\rm tot}^z=0$ sector; it has
dynamical exponent $z_0=2.71 < z$. 
The proximity of $z_0$ to the previous DMRG result \cite{Dellanna16} suggests that these authors worked in the $S_{\rm tot}^z\!=\!0$
sector, thus missing the lowest excitations. We have also analyzed the excitations in the $S_{\rm tot}^z\!=\! 2$ sector and have found that they
have the same dynamical exponent as in the $S_{\rm tot}^z=1$ sector \cite{SM}.    
%%%%%%%%%%%%
%%%%%%%%%%%%

As we tune $c$ away from 1, the groundstate is still annihilated by all the local interaction terms, as is the case in \eqref{H} \cite{bravyi}. We find that $H$ remains gapless but that the dynamical exponent $z$ varies continuously with $c$. \rfig{z} shows that $z$ decreases monotonically as $c$ decreases, and $z>2$ except when $c=0$.  
Our DMRG results thus rule out the field theory above as the correct low energy description 
of $H$ for $c>0$. Interestingly, it provides a concrete instance where different Hamiltonians, here the generalized Motzkin Hamiltonian Eq.\eqref{H}
and $H_{\rm orb}$ Eq.\eqref{orbi}, 
can have the same groundstate but markedly distinct excitations.   
At the special point $c=0$, $H$ has $z=2$ by virtue to a mapping to the ferromagnetic Heisenberg spin-1/2 chain \cite{bravyi}. 
In that case, the groundstate manifold becomes highly degenerate, {\color{black} growing exponentially with $N$}, and contains the all-zero product state.
%the product state $|00\cdots\rangle$ as the simplest groundstate.
We can construct an exact excited state in the form of a spin wave.
Since we are interested in the thermodynamic limit, we can work with an infinitely long chain, in which case 
the excited state reads:
$\sum_j e^{ikj} |u\rangle_j |0\rangle_{\rm rest}$, where $k$ specifies the wave number of the mode.  
The wave has $S_{\rm tot}^z=1$ and energy $1-\cos k$, leading to $z=2$ at small $k$.        

We now provide physical insight into the result $z>2$ observed when $c>0$. 
Earlier we have seen how the groundstate properties of the generalized Motzkin Hamiltonian map to the classical Brownian
motion of a particle. %Due to the special form of the Hamiltonian \eqref{H}, 
We can go further and study the \emph{full spectrum} of the Hamiltonian of Eq.\eqref{H}
by examining the non-equilibrium dynamics of a \emph{classical} 1d spin chain. Indeed, the RK form of $H$ ensures that
we can map the quantum spin chain problem to the dynamics of classical chain governed by a Markovian master equation \cite{Henley04,Moessner2001,Castelnovo2005}. The non-diagonal elements of the rate matrix $W$ are given by the matrix elements of $H$ between spin configurations, $W_{C,C'}=-\langle C|H|C' \rangle$; the diagonal elements of $W$ follow from detailed balance.   
The quantum dynamics of \eqref{H} thus maps to the critical slowing down of the corresponding classical model endowed with dynamics $W$. 
%which can be studied via classical Monte Carlo simulations. 

Hohenberg and Halperin have classified the critical slowing down of classical critical models according to the symmetries 
of the low energy modes \cite{Hohenberg1977}. For instance, the Glauber dynamics of an Ising chain belongs to Model A because 
single spin flips (non-conserving) processes are allowed.   
In this case, the motion of a domain wall is described by the diffusion equation yielding a dynamical exponent $z=2$ \cite{Hohenberg1977}. 
In contrast, the generalized Motzkin Hamiltonian \eqref{H} maps to classical dynamics described by the Model B universality class since  
the spin flipping processes preserve $S^z_{\rm tot}$. The conservation law constrains the spin flips and can thus  
slow down the dynamics, leading to a larger $z$ \cite{Grynberg,Isakov2011}.  
For instance, the spin-conserving Kawasaki dynamics of an Ising chain show $z\!\approx\! 3$ in a 
certain temperature range \cite{Cordery1981,Grynberg}.  
In \rfig{z}, we also observe such subdiffusive behavior with $z\!\approx\! 3$.  
To understand such behavior, it is useful to analyze the physics near $c=0$, where we have shown above that  
$z=2$ results from the diffusive motion of an up spin in a sea of 0s. 
As we turn on $c>0$, the $|V\rangle\langle V|$ projector in Eq.\eqref{Hbulk} generates
more $ud$ pairs, which slow down the motion of the up spin. Indeed, imagine we create an $ud$ pair next to an $u$ spin, $0uud0$.
Using the projectors in Eq.\eqref{Hbulk}, it takes 3 moves to translate the leftmost $u$ one site to the right, as opposed to 1 move in the absence of the $ud$ pair.  
This argument suggests that the dynamics slow down as $c$ increases, in agreement with our DMRG results \rfig{z}, but doesn't explain the 
specific change of $z$ with $c$, nor the presence of excitations with different dynamical exponents. 
One can get quantitative results for $z$ by performing Monte Carlo sampling of the non-equilibrium classical spin chain, as we mention below. 
Finally, it would be desirable to obtain a field theory description for the full spectrum. We expect that marginal operators
play a key role in explaining the change of the dynamical exponents with $c$.     

{\bf Summary \& outlook.} We have studied the intricate dynamics of a $S=1$ quantum spin chain, Eq.\eqref{H}. 
Our DMRG simulations have revealed that the gapless system has a dynamical exponent $z$ that changes as a function of a coupling $c$
in the Hamiltonian, \rfig{z}.
Interestingly, similar behavior was observed in quantum critical lattice models in 2 spatial dimensions \cite{Isakov2011}, where the Hamiltonian also takes a RK form. 
These authors used the mapping to the non-equilibrium dynamics of a classical model described above in order to 
determine $z$ via classical Monte Carlo simulations; dynamics with a varying $z>2$ were also observed. It would be interesting to analyze these 2d Hamiltonians
further to see if an additional excitation with a smaller $z$ appears, just as in our case. 
The Monte Carlo methods could also be applied in our case to 
reach bigger system sizes.
 A further connection is that the 2d groundstates studied in \cite{Isakov2011}     
have an emergent \emph{spatial} conformal symmetry \cite{Isakov2011,Ardonne-2004,Fradkin-book}, a feature that also arises in the generalized Motzkin model, although in a more subtle way \cite{long-prep}. We should also note that systems with modes that scale with different dynamic critical exponents have been found in theories of the quantum nematic transition in 2d metals \cite{Oganesyan-2001,Meng2012,Lederer-2016}. 
Finally, although we focused on spin 1,
the physics we discussed applies to other quantum spin chains, such as spin $1/2$ ones, even without an RK structure \cite{long-prep}.
We thus see the emergence of a unified picture for a broad class of quantum critical systems with non-trivial dynamics. 
An important missing element in both 1d and 2d 
is a field theory description, although our present analysis might help guide the search. 
This program will also shed light on the non-equilibrium dynamics of classical systems via the exact map of Henley discussed above \cite{Henley04}.  

%\emph{Acknowledgments---}
\begin{acknowledgments}
We thank A.~Ludwig, R.~Movassagh, S.~Sachdev, M.~Stoudenmire and X.~Yu for useful discussions.  
XC was supported by a postdoctoral fellowship from the Gordon and Betty Moore Foundation,
under the EPiQS initiative, Grant GBMF4304, at the Kavli Institute for Theoretical
Physics. 
This work was supported in part by the US  National Science Foundation through grant DMR 1408713 at the University of Illinois (EF).
WWK was funded by a Discovery Grant from NSERC, and by a Canada Research Chair. 
The DMRG simulations were performed using the ITensor package (v2). We acknowledge support from the Center for Scientific Computing from the CNSI, MRL: an NSF MRSEC (DMR-1121053).
\end{acknowledgments} 

\bibliography{biblio}

\pagebreak
\newpage
\appendix
\begin{center}
{\Large {\bf Supplemental Material}}
\end{center}

\section{Generalized Motzkin Hamiltonian in spin language}
\label{ap:spin}

The full Hamiltonian for a chain of $N$ sites reads 
\begin{align}
  H = H_{\rm bulk}(c) + H_{\rm bdy}
\end{align}
where the bulk Hamiltonian is a sum of $D,U,V$ projectors:
\begin{align}
  H_{\rm bulk}(c) &=\sum_{i=1}^{N-1}\Pi_{i,i+1}\\
&= \sum_{i=1}^{N-1} \ket{D}_{\!i,i+1}\!\!\bra{D} + \ket{U}_{\!i,i+1}\!\!\bra{U}+ c \ket{V}_{\!i,i+1}\!\!\bra{V} \nonumber
\end{align}
with $\ket{D}= \tfrac{1}{\sqrt 2}( \ket{0d} -\ket{d0})$, $\ket{U}= \tfrac{1}{\sqrt 2}( \ket{0u} -\ket{u0})$,
$\ket{V}= \tfrac{1}{\sqrt 2}( \ket{00} -\ket{ud})$ and the parameter $c\geq0$. Here, $u,d,0$ %$\{u\!=\!\mathrm{up},d\!=\!\mathrm{down},0\}$ 
label the $S^z$ eigenstates. 
These projectors can be written in terms of spin operators: 
\begin{align}
4|D\rangle_{1,2}\langle D|=& S_1^z+S_2^z+ (S_1^z)^2+(S_2^z)^2\nonumber\\
&-(S_1^z)^2S_2^z-S_1^z(S_2^z)^2-2(S_1^z)^2(S_2^z)^2\nonumber\\
&-S_1^-S_1^zS_2^zS_2^+ - S_1^z S_1^+S_2^-S_2^z
\end{align}
\begin{align}
4|U\rangle_{1,2}\langle U|=& -S_1^z-S_2^z+ (S_1^z)^2+(S_2^z)^2\nonumber\\
& +(S_1^z)^2S_2^z +S_1^z(S_2^z)^2-2(S_1^z)^2(S_2^z)^2\nonumber\\
&-S_1^+S_1^zS_2^zS_2^- - S_1^z S_1^-S_2^+S_2^z
\end{align}
\begin{align}
4|V\rangle_{1,2}\langle V|=& 2-2(S_2^z)^2-2(S_1^z)^2-\frac{1}{2}S_1^zS_2^z\nonumber\\
&-\frac{1}{2} (S_1^z)^2S_2^z+\frac{1}{2}S_1^z(S_2^z)^2+ \frac{5}{2}(S_1^z)^2(S_2^z)^2\nonumber\\
&+S_1^-S_1^zS_2^+S_2^z + S_1^z S_1^+S_2^zS_2^-
\end{align}
where the raising/lowering operators are defined as usual:
\begin{align}
  S^\pm = S^x \pm i S^y
\end{align}
with the standard commutation relations
\begin{align}
  [S^\pm , S^z] = \mp S^\pm\,, \qquad  [S^+,S^-] = 2S^z
\end{align}
At $c=1$, the spin representation was also discussed in Ref.~\onlinecite{Movassagh2017}. 
We note that some terms in the $U,D,V$ projectors involve 3 spin operators. These can be converted to a 4-spin
interaction using the above commutation relations. % in order to get a bilinear-biquadratic Hamiltonian, as advertized in the main text.
We thus see that the Hamiltonian is of the 
bilinear-biquadratic form, $\sum_j (A_{ab} S_j^a S_{j+1}^b + B_{ab;cd} S_j^a S_j^b S_{j+1}^c S_{j+1}^d)$. 
This representation is convenient for the DMRG calculations. 

One can verify that neighboring $\Pi_{i,i+1}$ operators do not commute, e.g.\ 
\begin{empheq}{align}
  [\Pi_{12},\Pi_{23}]\neq 0 
\end{empheq}
However, the Hamiltonian is ``frustration-free'' in the quantum information sense of the word, because
all the $\Pi_{i,i+1}$ have at least one common zero-eigenvalue eigenstate:
\begin{align}
  \Pi_{i,i+1}\ket{\Psi_0}=0
\end{align}

It can be easily verified that the bulk Hamiltonian has a global $U(1)$ symmetry, i.e.\ we have the following commutation relation: 
\begin{align}
  \left[\sum_i S_i^z,H_{\rm bulk} \right]= 0
\end{align}
The boundary term is
\begin{empheq}{align} \label{eq:bdy}
  H_{\rm bdy} &= H_{\rm bdy}^{i=1} + H_{\rm bdy}^{i=N}  \\
  &= \frac12 S_1^z(S_1^z -1)+\frac12 S_N^z(S_N^z+1) 
\end{empheq}
It is important to realize that 
$H_{\rm bdy}$ does not commute with the bulk Hamiltonian:
\begin{align}
  [H_{\rm bdy},H_{\rm bulk}] \neq 0
\end{align}
The boundary term commutes with the symmetry generator $\sum_{i=1}^N S_i^z$, implying that the full Hamiltonian
is symmetric:
\begin{empheq}{align}
  \left[\sum_i S_i^z,H \right]= 0
\end{empheq}  

\section{Field theory calculations}   
\label{ap:qft}
For the groundstate wavefunction defined in Eq.(5) of the main text, %\eqref{psi0}, 
if we change $\phi\to X$ and $x\to\tau$, 
we recognize that it is the path integral for a quantum mechanical particle in imaginary time $\tau$. 
The particle's Hamiltonian in the position basis is ($\hbar=1$)
\begin{align}
  \hat H_1 = -\frac{1}{4\kappa} \frac{d^2}{dX^2} + V(X)
\end{align}
with mass $m=2\kappa$. The potential is: 
\begin{align}
  V(X<0) = \infty\,, \qquad V(X\geq 0) = 0
\end{align}
The infinite wall for $X<0$ prevents the particle from penetrating the negative half-line, 
i.e.\ it enforces the ``orbifold condition'' on the $\phi$ field. 
The eigenstates of $\hat H_1$ are forced to vanish beyond the wall $X<0$, but are otherwise
eigenstates of the free particle Hamiltonian when $X>0$. We thus find the eigenfunctions 
$\psi_k(X)=\sqrt{2/\pi}\sin(kX)$ for $X>0$ and zero when $X<0$, with $k\geq 0$ being a continuous wave
number. The corresponding energies are $E_k=k^2/(2m) = k^2/(4\kappa)$.  

Based on this complete set of eigenfunctions, we can evaluate the propagator $G$:
\begin{align}
&  G(X_f,\tau_f;X_i,\tau_i) \equiv\bra{X_f} e^{-(\tau_f-\tau_i)\hat H_1} \ket{X_i}\nonumber\\
&= \int_0^\infty dk\, \bra{X_f} e^{-(\tau_f-\tau_i)\hat H_1}\ket{\psi_k}\langle{\psi_k}|X_i\rangle\nonumber  \\
  &= \sqrt{\frac{\kappa}{\pi\,(\tau_f-\tau_i)}}\left[ e^{-\frac{\kappa(X_f-X_i)^2}{ (\tau_f-\tau_i)}} - e^{-\frac{\kappa(X_f+X_i)^2}{ (\tau_f-\tau_i)}} \right]
  \label{eq:propag-constr}
\end{align}
Notice there's no $i$ in the evolution operator because we work in Euclidean time. 
From the propagator $G$, we can further calculate the probability distribution function for the particle 
being at $X$ at time $\tau$:
\begin{align} \label{eq:X-distr}
  f(X,\tau) = \frac{\langle X_f,\tau_f|X,\tau\rangle \langle X,\tau|X_i,\tau_i\rangle}{\langle X_f,\tau_f|X_i,\tau_i\rangle} 
\end{align}
Going back to the original wavefunction, %in Eq.\eqref{psi0}, 
this gives the probability function for being at height $\phi$ at position $x$. We impose the boundary conditions
\begin{align}
  X_i=X_f=\delta, \quad \tau_i=0,\quad \tau_f=L
\end{align}
where we have introduced a regulator $\delta=0^+$. Using Eq.\eqref{eq:propag-constr}, we then find for $\phi\geq 0$
\begin{align}  \label{eq:f-orb}
  f(\phi,x) = \frac{1}{2\sqrt{\pi}} \left(\frac{4\kappa L}{ (L-x)x}\right)^{\!3/2} \! \phi^2\exp\left[\frac{-\kappa L \phi^2}{ (L-x)x} \right] 
\end{align}
where we have changed notation back, i.e., $X\to\phi$ and $\tau\to x$, as suitable for the height 
quantum field $\phi$. We note that the factor of $\phi^2$ results from the constraint $\phi>0$, which makes the distribution
function $f$ deviate from a Gaussian.
One can verify that the probability distribution is properly normalized: $\int_0^\infty d\phi \, f(\phi,x)=1$. 

%To obtain the 1-point function of the height field, we simply evaluate
%\begin{empheq}{align}
%  \langle \phi(x)\rangle &= \int_0^\infty d\phi\; \phi f(\phi,x) \\
%  &= 4\sqrt{\frac{\kappa x(L-x)}{\pi\, L} }
%\end{empheq}
%We also give the expectation value of $\phi^2$:
%\begin{empheq}[box=\fbox]{align}
%  \langle \phi(x)^2\rangle = \frac{6\kappa (L-x)x}{L}
%\end{empheq}
%Notice that both expectation values vanish at $x=0$ and $L$, as required by the Dirichlet boundary conditions. 
%%Also, the are well-defined in the thermodynamic limit $L\to\infty$, 
%%with $x>0$ held finite: $\langle \phi(x)\rangle=(2^{3/2}/\pi) \sqrt x$ and $\langle\phi(x)^2\rangle= 3x$. 
%These results naturally agree with the scaling dimension of $\phi$. 

Using the probability distribution \req{f-orb}, we can evaluate the R\'enyi entanglement entropies for region $A$ being
the interval $[0,L_A]$, as given in Eq.(7)%\eqref{EE} 
of the main text. The reduced density matrix for region $A$ is  
\begin{align}
\rho_A=\sqrt{\epsilon}f(\sqrt{\epsilon}\varphi, L_A)|\varphi\rangle\langle\varphi|
\end{align}
where $\varphi=\phi/\sqrt\epsilon$ is the dimensionless height variable, normalized by the short distance cutoff $\epsilon$. For general $n$, the R\'enyi entropy is 
\begin{empheq}{align}
  S_n &\equiv\frac{1}{1-n} \ln \Tr(\rho_A^n)\nonumber\\
&=\frac12\ln\left[\frac{L_A(L-L_A)}{\epsilon\, L} \right] + b(n) \label{eq:Sn-orb}
\end{empheq} 
where $\epsilon$ is a short distance cutoff and 
\begin{align}
 b(n) =&- \frac{(2n+1)\ln n +n\ln(\pi/(16\kappa)) +\ln(4\kappa)}{2(1-n)}  \nonumber\\
& +\frac{\ln\Gamma(n+\tfrac12) }{1-n}
\end{align}
The leading log term for $S_n$ agrees with the result for the Motzkin spin Hamiltonian \cite{bravyi,Movassagh2017}. 
The second term also agrees with the lattice result if we choose $\kappa=3/4$, which is the value derived in the 
main text. In the limit  $L_A\ll L$, we obtain:
\begin{align}
  S_n = \frac12\ln\left( \frac{L_A}{\epsilon} \right) + \cdots
\end{align}
Thus the EE has a logarithmic dependence on $L_A$, suggesting that the groundstate is highly entangled \cite{bravyi}. 

\subsection{Mutual information}
\begin{figure}%[hbt]
\centering
\includegraphics[width=.47\textwidth]{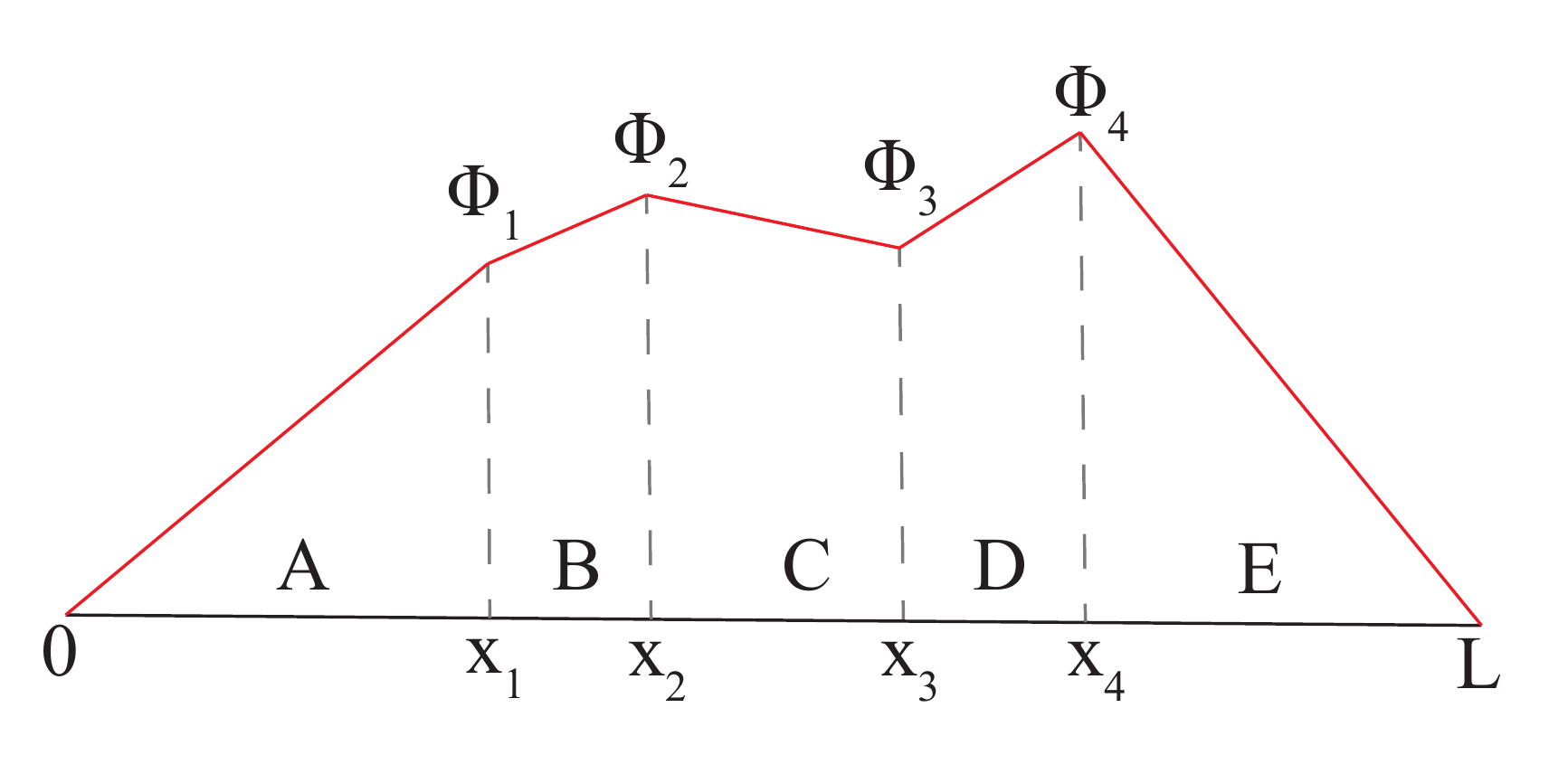}
\caption{A configuration with $\phi_i$ at $x_i$. }
\label{fig:mu_inf}
\end{figure}

We consider two short intervals $B$ and $D$ (Fig.~\ref{fig:mu_inf}) deep inside the bulk and compute the mutual information 
between them. If $L_B,L_C,L_D\ll L_A, L_E$, the probability for the height field $\phi$ in region $B, C, D$ to
approach zero is exponentially small.
Therefore the physics in the bulk region can be essentially described by the unconstrained $z=2$ boson.

%%%%%%%%%%%%%%
%%%%%%%%%%%%%% 

For the unconstrained boson, the probability distribution for the height field is simply a Gaussian distribution 
and it is easy to show that reduced 
density matrix for region $B\cup D$ depends only on $\phi_B\equiv\phi_2-\phi_1$, $\phi_D\equiv\phi_3-\phi_4$, where $\phi_i=\phi(x_i)$, 
and is equal to  
\begin{multline}
\rho_{B\cup D}\sim \int\! \mathcal D \phi_B \mathcal D \phi_D  \\
\times e^{-\tfrac{\kappa L_D(L-L_D)\phi_B^2+\kappa L_B(L-L_B)\phi_D^2-2\kappa L_BL_D\phi_B\phi_D}{L_BL_D(L-L_B-L_D)}}\\ 
\times |\phi_B,\phi_D\rangle\langle \phi_B,\phi_D|
\label{rho_BD}
\end{multline}
Therefore the R\'enyi entropy for $B\cup D$ is
\begin{align}
S_n(B\cup D)=&\frac{1}{2}\ln\left(\frac{L_BL_D}{\epsilon^2}\right)+\frac{1}{2}\ln\frac{L-L_B-L_D}{L}\nonumber\\
&+\ln(\pi/\kappa)-\frac{1}{(1-n)}\ln n
\end{align}
Similarly, we can also calculate EE for each interval and we have
\begin{align}
S_n(B)&=\frac{1}{2}\ln(\pi/\kappa)+\frac{1}{2}\ln \frac{L_B}{\epsilon}+\frac{1}{2}\ln\frac{L-L_B}{L}\nonumber\\
&-\frac{1}{2(1-n)}\ln n\nonumber\\
S_n(D)&=\frac{1}{2}\ln(\pi/\kappa)+\frac{1}{2}\ln \frac{L_D}{\epsilon}+\frac{1}{2}\ln\frac{L-L_D}{L}\nonumber\\
&-\frac{1}{2(1-n)}\ln n
\end{align}
The mutual information between $B$ and $D$ is
\begin{align}
I_n(B,D)&=S_n(B)+S_n(D)-S_n(B\cup D)\nonumber\\
&=\frac{1}{2}\ln\frac{(L-L_B)(L-L_D)}{(L-L_B-L_D)L}
\end{align}
We first note that $I_n(B,D)$ does not depend on the R\'enyi index. In the limit $L_B,L_D\ll L$, the mutual information vanishes.

% **************\\
% We note that Eq.\eqref{rho_BD} is invariant under the partial transpose operation for $\phi_D$ 
% and therefore the entanglement negativity between $B$ and $D$ is strictly equal to zero, as explained in the main text. 

\bl{\subsection{Logarithmic negativity}     
We now turn to the logarithmic negativity, which detects the entanglement between 2 disjoint
regions and is 
defined as 
\begin{align}
  \mc N\!=\! \ln \|\rho_{B\cup D}^{T_D}\|,   
\end{align} 
where $\rho_{B\cup D}^{T_D}$ denotes the partial transpose of the reduced density matrix $\rho_{B\cup D}$ with respect to region $D$. 
$\|\mc O \|$ is the sum of the absolute value of the eigenvalues of $\mc O$.  
For 2 intervals deep in the bulk, 
% \begin{align}
%   \rho_{B\cup D}\!=\! \int\! \mathcal D \phi_A\int\! \mathcal D \phi_B f(\phi_A,\phi_B)|\phi_A,\phi_B\rangle\langle \phi_A,\phi_B|
% \end{align}
$\rho_{B\cup D}$ given in Eq.\eqref{rho_BD} %is invariant under the partial transpose operation for $\phi_D$ 
is diagonal and invariant under the partial transpose for $\phi_D$, in contrast to the generic case.    
Thus $\mc N \!=\!\ln\mathrm{tr}(\rho_{B\cup D}) \!=\! 0$, further supporting the claim that the Motzkin wavefunction  
is less entangled than the groundstate of a CFT \cite{Calabrese2012}.}

\section{DMRG calculations}
\label{ap:dmrg}

We have numerically calculated the energy gap between the groundstate and the lowest energy excited states $\Delta E$ 
by using both exact diagonalization (ED) and the density matrix renormalization group (DMRG). 
The ED method is used for small systems $N\leq 10$ as a benchmark, 
and we perform large-scale DMRG calculations using the open-source C++ library ITensor. 

For the generalized Motzkin model, $\Delta E$ scales as $1/N^z$ for sufficiently large $N$. 
According to the previous ED result on small systems at $c=1$, the dynamical exponent $z$ is large and close to 3 
\cite{bravyi}. Meanwhile, this model is spin $S=1$. Both these factors make the DMRG more difficult, especially
if high precision is required. As a first step, we compare the groundstate energy and the von Neumann EE obtained via DMRG
with the analytical results and find that they agree precisely.
Then, in order to calculate the lowest excitations, which is doubly degenerate and has $S^z_{\rm tot}=\pm 1$, a large number of sweeps 
is used to ensure that the gap is well converged  
(the energy deviation for the last two sweeps is less than $10^{-12}$).  

\begin{figure}
\centering
\includegraphics[width=2.95in]{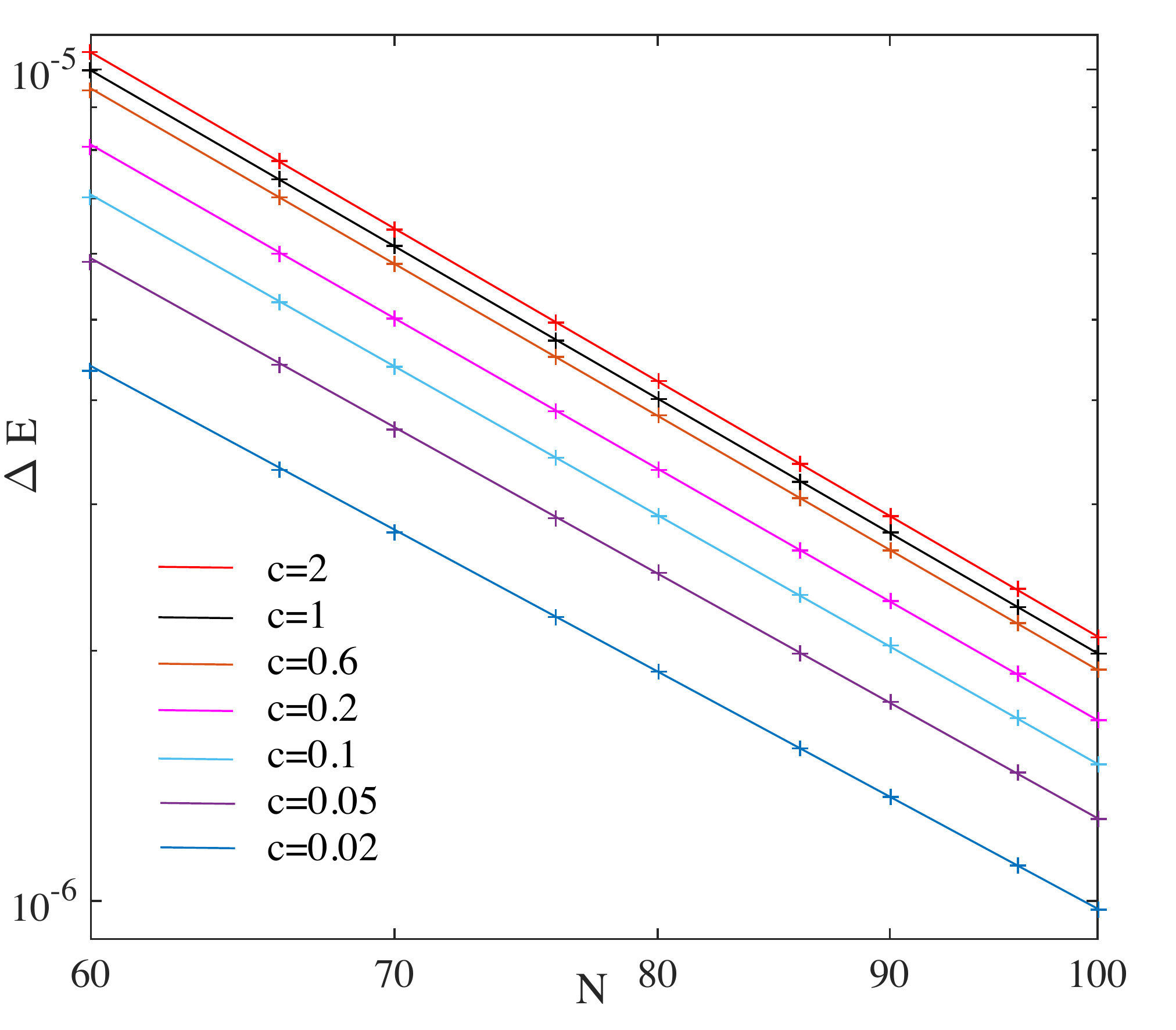}
\caption{Log-log plot of the energy gap $\Delta E$ versus system size $N$ for various $c$. The lines are fits to 
$\Delta E\propto N^{-z}$ using the data points $N\geq 80$. } 
\label{fig:E_vs_N} 
\end{figure}

We calculate the energy for the lowest excited state in the $S^z_{\rm tot}=1$ sector. We show the energy gap $\Delta E$ between 
the groundstate and this state in Fig.~\ref{fig:E_vs_N} as a function of system size. 
We fit $\Delta E\propto 1/N^z$ using the data points in the range $80\leq N\leq 100$ in order to minimize finite-size effects.
When $N\geq 80$, $\Delta E$ is less than $10^{-5}$ and decreases as $c$ decreases, which makes the small $c$ simulations more difficult.

\begin{figure}
\centering
\includegraphics[width=2.95in]{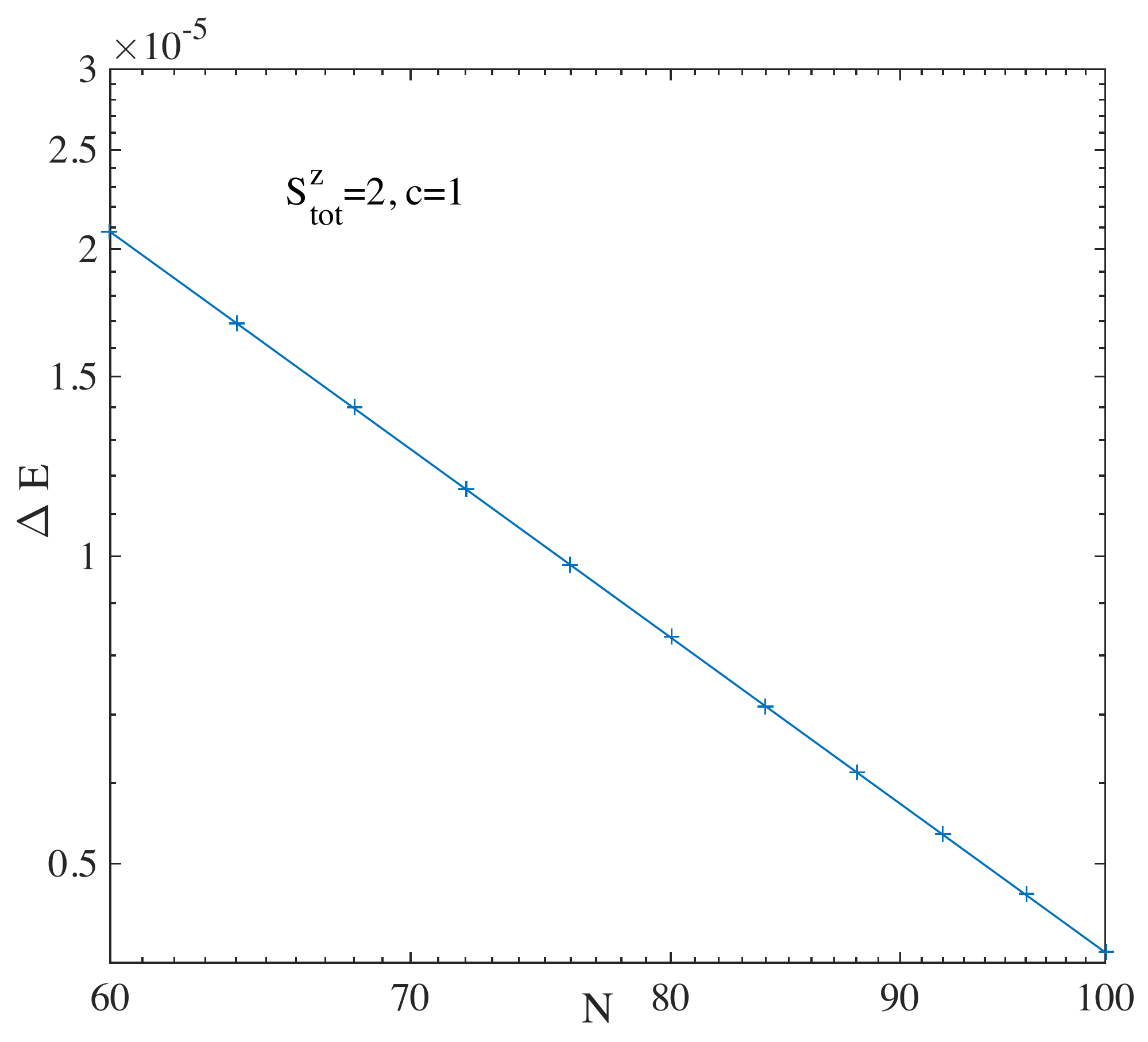}  
\caption{\bl{Log-log plot of the energy gap $\Delta E$ in the $S_{\rm tot}^z=2$ sector versus system size $N$ at $c=1$. The line is a fit to 
$\Delta E\propto N^{-z'}$ using the data points $N\geq 80$. The fit yields $z'=3.18$.} } 
\label{fig:Sz2}  
\end{figure}

The dynamical exponent $z$ is obtained by finite size scaling with $N$ from 80 to 100, with the detail is shown in Table \ref{table_z}. Moreover, we calculate the lowest excited state with $S^z_{\rm tot}=0$ and show the dynamical exponent $z_0$ in Table \ref{table_z}. 
We consider smaller system sizes $52\leq N\leq 80$ because 
the calculation of the excited state in the same spin sector as the groundstate 
is time-consuming. \bl{In Fig.~\ref{fig:Sz2}, we show the size dependence of the gap in the $S_{\rm tot}^z=2$ sector. The fit 
yields a dynamical exponent of $z'=3.18$, which is very close to the value obtained in the $S_{\rm tot}^z =1$ sector, $z=3.16$. 
This suggests that both sectors have the same dynamical exponent.
} 

%%%%%%%%%%%%%%
%%%%%%%%%%%%%

%%%%%%%%%%%%%%
%%%%%%%%%%%%%

%%%%%%%%%%%%%%
\begin{figure}
\centering
\includegraphics[width=2.95in]{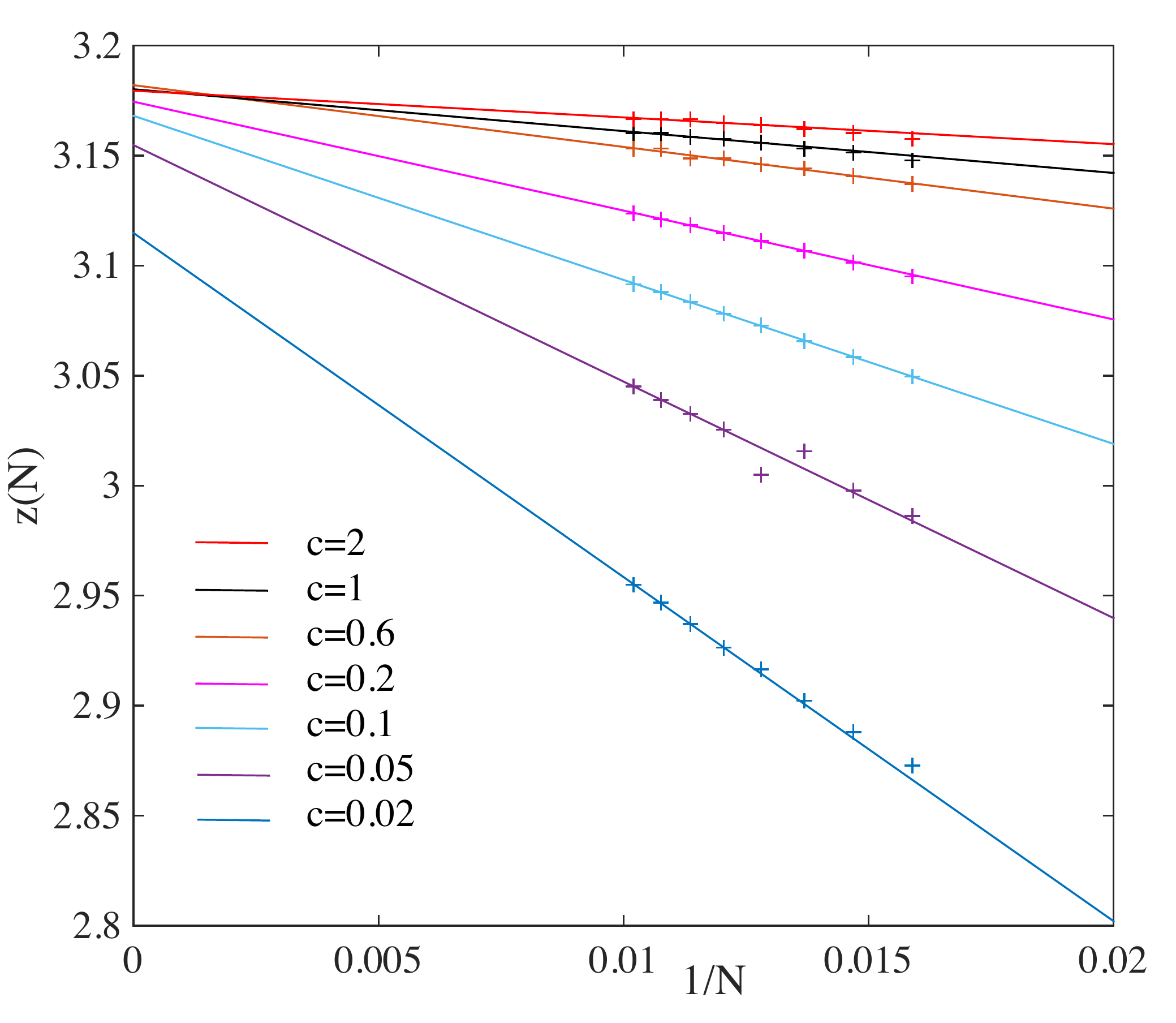}
\caption{Dynamical exponent $z(N)$  versus $1/N$ for various $c$. The crosspoint with vertical axis is $z(\infty)$.} 
\label{fig:z_linear} 
\end{figure} 
%%%%%%%%%%%%%

\begin{table}%[htbp]
\centering
\begin{tabular}{c||c|c|c|c}
$c$ & $z$ & $z(\infty)$  & $z_0$ & $z_0(\infty)$ \\ \hline  \hline
0 & $2$ &  & 2&\\
0.02 & $2.94$  & $3.11$ &  \\
0.05 & $3.03$ & $3.15$ & \\
0.1 & $3.08$ & $3.17$ &  $2.63$ & $2.70$\\
0.2 & $3.12$ &$3.17$ &  \\
0.4 & & & $2.70$ & $2.70$\\
0.6  & $3.15$ & $3.18$ &   \\
1 & $3.16$ & $3.18$ & $2.71$ & $2.70$\\
2 & $3.17$ & $3.18$ & \\ 
\end{tabular}
\caption{The dynamical exponents of the generalized Motzkin spin chain as a function of $c$, both in the $S^z_{\rm tot}=1$ and $S^z_{\rm tot}=0$ sectors. The lowest lying excited state has $S^z_{\rm tot}\!=\! 1$. The $z(\infty)$ results are obtained
using the extrapolation in \rfig{z_linear}.}\label{table_z} 
\end{table}

In \rfig{z_linear}, we show an alternate method to estimate $z$ in the thermodynamic limit. 
For 2 consecutive values of $N$, $N_1<N_2$, we evaluate the finite size exponent at the midpoint:
\begin{align}
  z \!\left(\frac{N_1+N_2}{2}\right) = - \frac{\ln(\Delta E_2/\Delta E_1)}{\ln (N_2/N_1)}  
\end{align}
The dependence of $z(N)$ on $N$ is shown in Fig.~\ref{fig:z_linear} for different values of the coupling $c$. 
We notice that the variation of $z(N)$ with $N$ is small when $c\geq 0.6$. As we decrease $c$, finite size effects for $z(N)$ 
become larger and we notice that $z(N)$ roughly scales as $1/N$. In order to get an estimate for $z$ in the thermodynamic limit,
we use a $1/N$ fit for the 4 largest $N$ points ($N>80$) to extract $z(N\to\infty)$. 
We compare the corresponding results with $z$ obtained from the fits described above 
in Table \ref{table_z}. 
It is important to emphasize that the $1/N$ extrapolation for $z(N)$ may not be accurate at $c<0.6$, and is only an estimate for the true $z$.
For instance, the slope could change at larger $N$, or deviate from $1/N$ scaling.

\end{document}